\documentclass[12pt]{article}
 
\usepackage{amsmath,amssymb}
\usepackage{bm}
\usepackage{graphicx, color}
\usepackage{wrapfig}
\usepackage{cite}

\usepackage{indentfirst}

\begin{document}
\title{
\begin{flushright}
\ \\*[-80pt] 
\begin{minipage}{0.22\linewidth}
\normalsize
EPHOU-21-003
UME-PP-017
 \\*[50pt]
\end{minipage}
\end{flushright}
{\Large \bf 
	Soft supersymmetry breaking terms and lepton flavor violations
	in modular flavor  models
	\\*[20pt]}}

\author{ 
\centerline{
Tatsuo Kobayashi $^{1}$,  Takashi Shimomura  $^{2}$ and 
 Morimitsu Tanimoto  $^{3}$
} 
\\*[20pt]
\centerline{
\begin{minipage}{\linewidth}
\begin{center}
$^1${\it \normalsize
Department of Physics, Hokkaido University, Sapporo 060-0810, Japan} \\*[5pt]
$^2${\it \normalsize
Faculty of Education, Miyazaki University, Miyazaki, 889-2192, Japan} \\*[5pt]
$^3${\it \normalsize
Department of Physics, Niigata University, Niigata 950-2181, Japan}
\end{center}
\end{minipage}}
\\*[50pt]}
\date{
\centerline{\small \bf Abstract}
\begin{minipage}{0.9\linewidth}
\medskip 
\medskip 
\small
\mbox{}
We study the soft supersymmetry (SUSY) breaking terms due to the modulus F-term  
in the modular flavor models of leptons.  It is found that the soft SUSY breaking terms
are constrained by the modular forms, and 
specific patterns are derived.
Those phenomenological implications are discussed in such as 
the lepton flavor violation  $\mu \rightarrow e + \gamma$ and $\mu \to 3e$ decays and $\mu \to e$ conversion 
in nuclei.
In order to examine numerically,  two modular flavor $A_4$ models
are taken.
The SUSY breaking scale is significantly constrained by inputting the observed upper bound of 
the $\mu \rightarrow e + \gamma$ decay.
The SUSY mass scale  is  larger than around $8$\,TeV
and $5$\,TeV for the two $A_4$ models, respectively.
Therefore, the  current experimental upper bound for the $\mu \to e + \gamma$ decay
corresponds to  the new physics of  the SUSY particle at the  $5$\,--\,$10$\,TeV scale  in the modular flavor  models.
The SUSY scale will be explored by future experiments of lepton flavor violation up to $8$\,--\,$17$ TeV.
The  predicted branching ratio depends on a modulus $\tau$ significantly.
It decreases of one order  at the large  ${\rm Im}\,\tau$.
We  also calculate the branching ratios of tauon decays to  $e + \gamma$ and $\mu + \gamma$.
Predicted ones  are
at most ${\cal O}(10^{-15})$, which are much  below the current experimental  bounds. 
\end{minipage}
}

\begin{titlepage}
\maketitle
\thispagestyle{empty}
\end{titlepage}

\section{Introduction}
\mbox{}

The origin of flavor  is one of the important issues in particle physics.
Non-Abelian flavor symmetries are interesting approaches among various 
approaches to understand the flavor origin.
Indeed, a lot of  works have been presented by using various non-Abelian discrete  groups for flavors to understand the flavor structures of quarks and leptons.
Those are 
motivated by the precise observation of  flavor mixing angles of  leptons
\cite{Altarelli:2010gt,Ishimori:2010au,Ishimori:2012zz,Hernandez:2012ra,King:2013eh,King:2014nza,Tanimoto:2015nfa,King:2017guk,Petcov:2017ggy,Feruglio:2019ktm}.
In particular, the $A_4$ flavor models are attractive  
because the $A_4$ group is the minimal one including a triplet 
irreducible representation, 
which allows for a natural explanation of the  
existence of  three families of quarks and leptons 
\cite{Ma:2001dn,Babu:2002dz,Altarelli:2005yp,Altarelli:2005yx,
	Shimizu:2011xg,Petcov:2018snn,Kang:2018txu}.
In spite of such a theoretical effort,  we have not yet the fundamental theory of flavor.

Flavor symmetries control not only the flavor structure of quarks and leptons, but also 
the flavor structure of their superpartners and lead to specific patterns in soft 
supersymmetry (SUSY) breaking terms.
Soft SUSY breaking terms were studied in several models with 
non-Abelian flavor symmetries \cite{Ko:2007dz,Ishimori:2008ns,Ishimori:2008au,Ishimori:2009ew}, 
and they are different from patterns of soft SUSY breaking terms 
in other flavor models. (See e.g. \cite{Kobayashi:2000br}.)
Such structure can be observed directly and/or indirectly if 
the mass scale of superpartners is light enough.
For example, flavor changing processes are important to test 
the flavor structure of superpartners different from the flavor structure 
of quarks and leptons.

A new direction to flavor symmetry,
 modular invariance has been proposed  in the lepton sector \cite{Feruglio:2017spp}.
The modular symmetry arises from the compactification of a higher dimensional theory on a torus or an orbifold as well as low-energy effective field theory of superstring theory \cite{Lauer:1989ax,Lerche:1989cs,Ferrara:1989qb,Kobayashi:2016ovu,Kobayashi:2018rad,Kikuchi:2020frp}.
The shape of the compact space is parametrized by a modulus $\tau$ living in the upper-half complex plane, up to modular transformations.
The finite groups $S_3$, $A_4$, $S_4$, and $A_5$
are isomorphic to the finite modular groups 
$\Gamma_N$ for $N=2,3,4,5$, respectively\cite{deAdelhartToorop:2011re}.

In this approach, fermion matrices are written in terms of modular forms which are   holomorphic functions of  the modulus  $\tau$.
The lepton mass matrices have given successfully  in terms of  $A_4$ modular forms \cite{Feruglio:2017spp}.
Modular invariant flavor models have been also proposed on the $\Gamma_2\simeq S_3$ \cite{Kobayashi:2018vbk},
$\Gamma_4 \simeq S_4$ \cite{Penedo:2018nmg} and  
$\Gamma_5 \simeq A_5$ \cite{Novichkov:2018nkm}.
Based on these modular forms, flavor mixing of quarks/leptons have been discussed intensively in these years.

The vacuum expectation value (VEV) of the modulus $\tau$ plays a role in modular flavor 
symmetric models, in particular realization of quark and lepton masses and 
their mixing angles.
The modulus VEV is fixed as the potential minimum of the modulus potential.
(See for the modulus stabilization in modular flavor models, e.g. \cite{Kobayashi:2019xvz,Kobayashi:2019uyt,Kobayashi:2020uaj,Ishiguro:2020tmo}.)
At such a minimum, the F-term of the modulus $F^\tau$ may be non-vanishing, 
and lead to SUSY breaking, 
the so-called moduli-mediated SUSY breaking \cite{Kaplunovsky:1993rd,Brignole:1993dj,Kobayashi:1994eh,Ibanez:1998rf}, although there may be other sources of SUSY breaking.
That leads to specific patterns of soft SUSY breaking terms.
Thus, our purpose in this paper is to study such specific patterns of soft SUSY breaking terms due to 
$F^\tau$ and its phenomenological implications such as the lepton flavor violations.\footnote{
Recently,  in Ref.~ \cite{Du:2020ylx},   SUSY breaking phenomenology 
was studied in the modular flavor $S_3$ invariant SU(5) GUT model \cite{Kobayashi:2019rzp}
by assuming the F-term of {\bf 24} chiral field.}



We study  the soft SUSY breaking terms  in the modular flavor models of leptons.  It is found that the soft SUSY breaking terms
are constrained by the modular forms and there appears a specific pattern of 
soft SUSY breaking terms due to the modulus F-term 
in the modular flavor symmetric models.
In order to discuss the  soft SUSY breaking terms in  the lepton flavor violation (LFV),
we examine  numerically $\mu \rightarrow e + \gamma$ and $\mu \to 3e$ decays and $\mu \to e$ conversion 
in nuclei in the modular flavor  $A_4$ model.
The SUSY breaking scale is significantly constrained by inputting the observed upper bound of the $\mu \rightarrow e + \gamma$ decay 
\cite{TheMEG:2016wtm}.


In section 2,
we give a brief review on the modular symmetry. 
In section 3, we present the soft SUSY breaking terms in the modular flavor models.
In section 4, we calculate LFV,  e.g.,
the  $\mu \rightarrow e + \gamma$ and $\mu \to 3e$ decays and $\mu \to e$ conversion 
in nuclei in terms of the soft SUSY breaking masses in the modular flavor  $A_4$ models,
 and present numerical discussions.
Section 5 is devoted to a summary.
In Appendix A, the tensor product  of the $A_4$ group is presented.

\section{Modular group and modular forms}

The modular group $\bar\Gamma$ is the group of linear fractional transformations
$\gamma$ acting on the modulus  $\tau$, 
belonging to the upper-half complex plane as:
\begin{equation}\label{eq:tau-SL2Z}
\tau \longrightarrow \gamma\tau= \frac{a\tau + b}{c \tau + d}\ ,~~
{\rm where}~~ a,b,c,d \in \mathbb{Z}~~ {\rm and }~~ ad-bc=1, 
~~ {\rm Im} [\tau]>0 ~ ,
\end{equation}
which is isomorphic to  $PSL(2,\mathbb{Z})=SL(2,\mathbb{Z})/\{\mathbb{I},-\mathbb{I}\}$ transformation.
This modular transformation is generated by $S$ and $T$, 
\begin{eqnarray}
S:\tau \longrightarrow -\frac{1}{\tau}\ , \qquad\qquad
T:\tau \longrightarrow \tau + 1\ ,
\label{symmetry}
\end{eqnarray}
which satisfy the following algebraic relations, 
\begin{equation}
S^2 =\mathbb{I}\ , \qquad (ST)^3 =\mathbb{I}\ .
\end{equation}

We introduce the series of groups $\Gamma(N)~ (N=1,2,3,\dots)$,
called principal congruence subgroups, defined by
\begin{align}
\begin{aligned}
\Gamma(N)= \left \{ 
\begin{pmatrix}
a & b  \\
c & d  
\end{pmatrix} \in SL(2,\mathbb{Z})~ ,
~~
\begin{pmatrix}
a & b  \\
c & d  
\end{pmatrix} =
\begin{pmatrix}
1 & 0  \\
0 & 1  
\end{pmatrix} ~~({\rm mod} N) \right \}
\end{aligned} .
\end{align}
For $N=2$, we define $\bar\Gamma(2)\equiv \Gamma(2)/\{\mathbb{I},-\mathbb{I}\}$.
Since the element $-\mathbb{I}$ does not belong to $\Gamma(N)$
for $N>2$, we have $\bar\Gamma(N)= \Gamma(N)$.
The quotient groups defined as
$\Gamma_N\equiv \bar \Gamma/\bar \Gamma(N)$
are  finite modular groups.
In this finite groups $\Gamma_N$, $T^N=\mathbb{I}$  is imposed.
The  groups $\Gamma_N$ with $N=2,3,4,5$ are isomorphic to
$S_3$, $A_4$, $S_4$ and $A_5$, respectively \cite{deAdelhartToorop:2011re}.

Modular forms of  level $N$ are 
holomorphic functions $f(\tau)$  transforming under 
$\Gamma(N)$ as:
\begin{equation}
f(\gamma\tau)= (c\tau+d)^kf(\tau)~, ~~ \gamma \in \Gamma(N)~ ,
\end{equation}
where $k$ is the so-called  modular weight.

The low-energy effective field theory derived from superstring theory has also the modular symmetry.
Under the modular transformation of Eq.(\ref{eq:tau-SL2Z}), chiral superfields $\phi$ 
transform as \cite{Ferrara:1989bc},
\begin{equation}
\phi_i\to(c\tau+d)^{-k_i}\rho_{ij}(\gamma)\phi_j,
\end{equation}
where  $-k_i$ is the modular weight and $\rho_{ij}(\gamma)$ denotes a unitary representation matrix of $\gamma\in \bar\Gamma$.

We study global supersymmetric models, e.g., the minimal supersymmetric standard model.
The superpotential, which is built from matter fields and modular forms, 
is assumed to be modular invariant, i.e., to have 
a vanishing modular weight. For given modular forms 
this can be achieved by assigning appropriate
weights to the matter superfields.

The kinetic terms  are  derived from a K\"ahler potential.
The K\"ahler potential of chiral matter fields $\phi_i$ with the modular weight $-k_i$ is given simply  by 
\begin{equation}
K^{\rm matter} = K_{i \bar i}|\phi_i|^2,\qquad K_{i \bar i}= \frac{1}{[i(\bar\tau - \tau)]^{k_i}} ,
\end{equation}
where the superfield and its scalar component are denoted by the same letter, and  $\bar\tau =\tau^*$ after taking the VEV.
Therefore, 
the canonical form of the kinetic terms  is obtained by 
changing the normalization.

The modular forms of weight $k$ span the linear space ${\cal M}_k(\Gamma{(N)})$.
For example, for $\Gamma_3\simeq A_4$, the dimension of the linear space 
${\cal M}_k(\Gamma{(3)})$ is $k+1$ \cite{Gunning:1962,Schoeneberg:1974,Koblitz:1984}, i.e., there are three linearly 
independent modular forms of the lowest non-trivial weight $2$.
These forms have been explicitly obtained \cite{Feruglio:2017spp} in terms of
the Dedekind eta-function $\eta(\tau)$: 
\begin{equation}
\eta(\tau) = q^{1/24} \prod_{n =1}^\infty (1-q^n)~, 
\quad\qquad  q= \exp \ (i 2 \pi  \tau )~,
\label{etafunc}
\end{equation}
%
where $\eta(\tau)$ is a  so called  modular form of weight~$1/2$. 
In what follows we will use the following basis of the 
$A_4$ generators  $S$ and $T$ in the triplet representation:
\begin{align}
\begin{aligned}
S=\frac{1}{3}
\begin{pmatrix}
-1 & 2 & 2 \\
2 &-1 & 2 \\
2 & 2 &-1
\end{pmatrix},
\end{aligned}
\qquad \qquad
\begin{aligned}
T=
\begin{pmatrix}
1 & 0& 0 \\
0 &\omega& 0 \\
0 & 0 & \omega^2
\end{pmatrix}, 
\end{aligned}
\label{STbase}
\end{align}
%
where $\omega=\exp (i\frac{2}{3}\pi)$ .
The  modular forms of weight 2 transforming
as a triplet of $A_4$ can be written in terms of 
$\eta(\tau)$ and its derivative \cite{Feruglio:2017spp}:
\begin{eqnarray} 
\label{eq:Y-A4}
Y_1 &=& \frac{i}{2\pi}\left( \frac{\eta'(\tau/3)}{\eta(\tau/3)}  +\frac{\eta'((\tau +1)/3)}{\eta((\tau+1)/3)}  
+\frac{\eta'((\tau +2)/3)}{\eta((\tau+2)/3)} - \frac{27\eta'(3\tau)}{\eta(3\tau)}  \right), \nonumber \\
Y_2 &=& \frac{-i}{\pi}\left( \frac{\eta'(\tau/3)}{\eta(\tau/3)}  +\omega^2\frac{\eta'((\tau +1)/3)}{\eta((\tau+1)/3)}  
+\omega \frac{\eta'((\tau +2)/3)}{\eta((\tau+2)/3)}  \right) , \label{Yi} \\ 
Y_3 &=& \frac{-i}{\pi}\left( \frac{\eta'(\tau/3)}{\eta(\tau/3)}  +\omega\frac{\eta'((\tau +1)/3)}{\eta((\tau+1)/3)}  
+\omega^2 \frac{\eta'((\tau +2)/3)}{\eta((\tau+2)/3)}  \right)\,.
\nonumber
\end{eqnarray}
%
The overall coefficient in Eq.\,(\ref{Yi}) is 
one possible choice.
It cannot be uniquely determined.
The triplet modular forms of weight 2
have the following  $q$-expansions:
\begin{align}
{ Y^{(2)}_{\bf 3}}
=\begin{pmatrix}Y_1\\Y_2\\Y_3\end{pmatrix}=
\begin{pmatrix}
1+12q+36q^2+12q^3+\dots \\
-6q^{1/3}(1+7q+8q^2+\dots) \\
-18q^{2/3}(1+2q+5q^2+\dots)\end{pmatrix}.
\label{Y(2)}
\end{align}
%
They also satisfy the constraint \cite{Feruglio:2017spp}:
\begin{align}
Y_2^2+2Y_1 Y_3=0~.
\label{condition}
\end{align}

The  modular forms of the  higher weight, $k=4,\,6,\,8 \dots$, can be obtained
by the $A_4$ tensor products of  the modular forms  with weight 2,
${ Y^{(2)}_{\bf 3}}$ 
as given in Appendix A.


\section{Soft SUSY breaking terms}

We study soft SUSY breaking terms due to the modus F-term within the 
framework of supergravity theory, using the unit $M_P=1$, where 
$M_P$ denotes the reduced Planck scale.
The  full K\"ahler potential is given as: 
\begin{eqnarray}
K & =& K_0(\tau,M)+
K^{\rm matter} \,, \nonumber \\
K_0(\tau,M) &=& -\ln(i(\bar \tau -\tau)) + K(M,\bar M)\,, 
\label{kahler}
\end{eqnarray}
where $M$ denotes moduli other than $\tau$.
The superpotential is given as:
\begin{eqnarray}
W= Y_{ijk}(\tau)\Phi_i \Phi_j \Phi_k  + M_{ij}(\tau)\Phi_i \Phi_j\cdots \,.
\label{super}
\end{eqnarray}
We assume that 
the gauge kinetic function is independent of the modulus $\tau$, i.e. $f(M)$.

Let us consider the case that the SUSY breaking is occurred 
by some F-terms of  moduli $X$, $F^X$ $(F^X\not= 0)$.
In the canonical normalization,
the soft masses $\tilde m_i$ and the A-term are given as \cite{Kaplunovsky:1993rd}:
\begin{eqnarray}
\tilde m_i^2= m_{3/2}^2-\sum_X |F^X|^2 \partial_X \partial_{\bar X}\ln K_{i \bar i}\,,
\end{eqnarray}
and 
\begin{eqnarray}
A_{ijk} =A_i+A_j+A_k -\sum_X\frac{F^X}{Y_{ijk}} \partial_X Y_{ijk}\,, \nonumber
\end{eqnarray}
\begin{eqnarray}
A_i = \sum_X F^X \partial_X \ln e^{-K_0/3}K_{i\bar i}\,,
\end{eqnarray}
where $i,\, j$ and $k$ denote flavors.
Here, Yukawa couplings $\tilde Y_{ijk}$ in global SUSY superpotential
are related with Yukawa couplings $ Y_{ijk}$ in the supergravity  superpotential as follows:
\begin{eqnarray}
|\tilde Y_{ijk}|^2=e^{K_0}|Y_{ijk}|^2\,.
\end{eqnarray}
That is, the global SUSY superpotential has vanishing 
modular weight, while  the supergravity  superpotential has 
the modular weight $-1$.
Some modular flavor models are studied in global SUSY basis.
At any rate, we can realize the same results of quark and lepton mass ratios and mixing angles 
by properly shifting assignment of modular weights for matter fields.

Suppose the case of   $X=\tau$. The K\"ahler potential $K$ in Eq.\,(\ref{kahler})
leads to  the soft mass
\begin{eqnarray}
\tilde m_i^2= m_{3/2}^2-k_i \frac{|F^\tau|^2}{(2{\rm Im}\tau)^2} \,.
\end{eqnarray}
The A-term is written by 
\begin{eqnarray}
A_{ijk}&=&A_{ijk}^0+A'_{ijk}, \nonumber \\
A_{ijk}^0&=& (1-k_i-k_j-k_k)\frac{F^\tau}{(2{\rm Im}(\tau))}, \qquad\qquad  A'_{ijk}=\frac{F^\tau}{Y_{ijk}}\frac{dY_{ijk}(\tau)}{d \tau} \, .
\label{Aterm}
\end{eqnarray}
Note that in our convention $\tau$ is dimensionless, and $F^\tau$ has the dimension one.
Gaugino masses can be generated by F-terms of other moduli, $F^M$, 
while $F^M$ have universal contributions on soft masses and A-terms.

Since  we have common weights for three generations in the simple modular flavor model,  the soft mass $\tilde m_i$ is flavor blind.
Therefore, we have universal mass matrices  
\begin{eqnarray}
\tilde m_{Li}^2=\tilde m^2_L,\qquad\qquad \tilde m^2_{ei}=\tilde m_e^2 \,,
\label{massLe}
\end{eqnarray}
that is, they are  proportional to the unit matrix.

The first term of $A_{ijk}$ term in Eq.\,(\ref{Aterm}), $A^0_{ijk}$ is also flavor blind.
If there is another source of SUSY breaking, $A^0_{ijk}$ is shifted by $\Delta A$ as
\begin{eqnarray}
A^0_{ijk} +\Delta A = A^0 \,,
\end{eqnarray}
where $\Delta A$ is also flavor blind.
Therefore, we write 
\begin{eqnarray}
A_{ijk}&=&A^0+A'_{ijk}\,, 
\label{aterm}
\end{eqnarray}
where the second term of r.h.s. in Eq.\,(\ref{aterm}) only depends on the flavor.
\section{A-term in modular $A_4$ flavor model}

We discuss the soft SUSY breaking terms in Eq.\,(\ref{aterm})
 in the modular $A_4$ models.
 In order to present the explicit form of the A-term,
 we consider successful lepton mass matrices to be consistent with
 observed lepton masses and flavor mixing angles.
A simple global SUSY model is shown in Table \ref{tb:lepton}, where  
the three left-handed lepton doublets $L$  compose a $A_4$ triplet,
and the right-handed charged leptons $e^c$, $\mu^c$ and $\tau^c$ are $A_4$ singlets. $H_u$ and $H_d$ are Higgs doublets.
The weight $k$'s of the superfields of left-handed leptons and 
right-handed charged leptons is $-2$ and $0$, respectively,
which are common for three generations.\footnote{We can construct the model, where 
the same modular forms appear in the Yukawa couplings and the Weinberg operators in the supergravity 
superpotential, by properly shifting the assignment of modular weights for matter fields. }
The charged lepton mass matrix is given 
in terms of  modular forms of $A_4$ triplet with weight $2$, 
${ Y_{\bf 3}^{\rm (2)}}$ simply.
The neutrinos are supposed  to be Majorana particles in Table \ref{tb:lepton}.
Since there are no  right-handed neutrinos,
neutrino mass matrix can be  written by using the Weinberg operator.
Then, the neutrino mass term is  given in terms of  modular forms of
 $A_4$ triplet ${Y_{\bf 3}^{\rm (4)}}$ and
  $A_4$ singlets ${ Y_1^{\rm (4)}}$ and  ${ Y_{1'}^{\rm (4)}}$
  with   weight $4$.
  This model has been discussed focusing on the flavor mixing
  numerically \cite{Okada:2019uoy,Okada:2020ukr,Okada:2020brs}.

\begin{table}[h]
	\centering
	\begin{tabular}{|c||c|c|c|c|c|} \hline
		\rule[14pt]{0pt}{1pt}
		&$L$&$(e^c,\mu^c,\tau^c)$&$H_u$&$H_d$&$Y_{\bf r}^{\rm (2)}, 
		\ \   Y_{\bf r}^{\rm (4)}$\\  \hline\hline 
		\rule[14pt]{0pt}{1pt}
		$SU(2)$&$\bf 2$&$\bf 1$&$\bf 2$&$\bf 2$&$\bf 1$\\
		\rule[14pt]{0pt}{1pt}
		$A_4$&$\bf 3$& \bf (1,\ 1$''$,\ 1$'$)&$\bf 1$&$\bf 1$&$\bf 3, \ \{3, 1, 1'\}$\\
		\rule[14pt]{0pt}{1pt}
		$k$&$ -2$&$(0,\ 0,\ 0)$ &0&0& \hskip -0.7 cm $2, \qquad 4$ \\ \hline
	\end{tabular}	
	\caption{ Representations of $SU(2)$, $A_4$ and  weights $k$ for matter fields and  modular forms of 
		    weight $2$ and $4$.  The subscript $\bf r$ represents the $A_4$ representation of modular forms. }
	\label{tb:lepton}
\end{table}

For charged lepton sector,
 Yukawa couplings $Y_{ijk}$ are given in terms of modular forms in Eq.\,(\ref{Y(2)})
\begin{align}
\begin{aligned}
Y_{ijk}&={\rm diag}[\alpha_e, \beta_e, \gamma_e]
\begin{pmatrix}
Y_1 & Y_3 & Y_2 \\
Y_2 & Y_1 & Y_3 \\
Y_3 & Y_2 & Y_1
\end{pmatrix}_{RL},
\end{aligned}\label{eq:CL}
\end{align}
where coefficients $\alpha_e$, $\beta_e$ and $\gamma_e$ are taken to be  real without loss of generality.


Since the A-term is given  in Eq.\,(\ref{aterm}),
 the soft mass term  $h_{ijk}=Y_{ijk}A_{ijk}$ is given 
\begin{align}
\begin{aligned}
h_{ijk}&=A_0\times {\rm diag}[\alpha_e, \beta_e, \gamma_e]
\begin{pmatrix}
Y_1 & Y_3 & Y_2 \\
Y_2 & Y_1 & Y_3 \\
Y_3 & Y_2 & Y_1
\end{pmatrix}_{RL} + F^\tau \times 
{\rm diag}[\alpha_e, \beta_e, \gamma_e]
\begin{pmatrix}
Y'_1 & Y'_3 & Y'_2 \\
Y'_2 & Y'_1 & Y'_3 \\
Y'_3 & Y'_2 & Y'_1
\end{pmatrix}_{RL},
\end{aligned}\label{eq:h}
\end{align}
where $Y'$ is the derivative of $Y$ with respect to $\tau$.

In the super-CKM (SCKM) basis, the first term of the r.h.s. 
in Eq.\,(\ref{eq:h}) is diagonal. Therefore,
the second term of the r.h.s. only contributes to the LFV.
Since the modulus  $\tau$ and couplings $\alpha_e, \beta_e, \gamma_e$
are fixed by the experimental data of neutrino oscillations
in the model of Table 1,
 we can estimate the magnitude of LFV if $F^\tau$ is given.

These expressions are given at high energy scale, for example, GUT scale.
The effects of the renormalization  group (RG) running on the soft mass terms should be taken into account
at the electroweak (EW) scale.
We consider the small $\tan \beta$ scenario, where  the Yukawa couplings of charged leptons and down-type quarks
are small.
Then, 
the largest contributions of the effect for off diagonal elements of the A-term are  those of gauge couplings. Then, we can estimate the running effects 
by 
\begin{eqnarray}
{A}_{ijk} (m_Z)
=\exp\left[ \frac{-1}{16\pi^2}\int_{m_Z}^{m_\text{GUT}} dt
~ \left ( \frac95 g_1^2+3g_2^2 \right )\right ]{A}_{ijk} (m_\text{GUT})
\approx 1.5\times {A}_{ijk} (m_\text{GUT}),
\end{eqnarray}
which is flavor independent.  Thus, the RG effect does not change
the flavor structure at the hight energy scale.
The RG effect can be absorbed in the averaged slepton mass
 at the low energy.
On the other hand, we do not need to discuss the A-term of the neutrino sector because there are no right-handed neutrinos.


\section{LFV in SUSY flavor violation}

The  SUSY flavor phenomena of LFV for the lepton sector were
discussed by introducing gauge singlet scalars (flavons) in  the non-Abelian discrete symmetry \cite{Feruglio,Ishimori:2010su}.
In contrast to previous works, our modular flavor $A_4$ models constrain 
the flavor changing processes  significantly via modular forms
as discussed in the previous section.

Let us define mass insertion parameters,  $\delta_\ell^{LL}$, $\delta_\ell^{LR}$,
$\delta_\ell^{RL}$ and   $\delta_\ell^{RR}$ by 
\begin{eqnarray}
m_{\tilde\ell}^2
\begin{pmatrix} \delta_\ell^{LL} & \delta_\ell^{LR} \\ 
\delta_\ell^{RL}    & \delta_\ell^{RR}  \\
\end{pmatrix}
=
\begin{pmatrix} \tilde m_{ L}^2 & \tilde m_{LR}^2 \\ 
\tilde m_{RL}^2    & \tilde m_{e}^2  \\
\end{pmatrix}
- \text{diag}(m_{\tilde\ell}^2) \ ,
\label{massinsertion}
\end{eqnarray}
where $m_{\tilde\ell}$ is an average slepton mass. 
Here, $ \tilde m_{L}^2$ and  $\tilde m_{R}^2$ are universal diagonal matrices as given in Eq.\,(\ref{massLe}).
By using $h_{ijk}=Y_{ijk}A_{ijk}$ in Eq.\,(\ref{eq:h}),
we get $\tilde m_{RL}^2= v_d h_{ijk}$, where $v_d$ is the VEV of the neutral component of Higgs doublet $H_d$.
We have also  $\tilde m_{LR}^2=\tilde m_{RL}^{2\ \dagger}$.

Let us examine the effect of the A-term on the LFV rare decay
such as $\ell_i\to\ell_j + \gamma$, $\ell_i \to \ell_j \ell_k \bar{\ell}_k$ and LFV conversion $\mu N \to e N$. 
Once non-vanishing off diagonal elements of the slepton mass matrices
are generated in the SCKM basis,
the LFV rare decays and conversion are naturally induced by one-loop diagrams with the exchange of gauginos 
and sleptons. 

The decay $\ell_i\to\ell_j + \gamma$ is described by the dipole operator and the corresponding amplitude reads
\cite{Borzumati:1986qx,Gabbiani:1996hi,Hisano:1995nq,Hisano:1995cp,	Hisano:2007cz,Hisano:2009ae,Altmannshofer}, 
\begin{eqnarray}
T=m_{\ell_i}\epsilon^{\lambda}\overline{u}_j(p-q)[iq^\nu\sigma_{\lambda\nu}
(A_{L}^{ij}P_{L}+A_{R}^{ij}P_{R})]u_i(p)\,,
\end{eqnarray}
where $p$ and $q$ are momenta of the initial lepton $\ell_i$
and of the photon, respectively, 
and $A_{L}^{ij},\,A_{R}^{ij}$ are the two possible amplitudes in this  process. 
The branching ratio of $\ell_{i}\rightarrow \ell_{j} +\gamma$ can be written 
as follows:
\begin{eqnarray}
\frac{{\rm BR}(\ell_{i}\rightarrow  \ell_{j}\gamma)}{{\rm BR}(\ell_{i}\rightarrow 
	\ell_{j}\nu_i\bar{\nu_j})} =
\frac{48\pi^{3}\alpha}{G_{F}^{2}}(|A_L^{ij}|^2+|A_R^{ij}|^2)\,,
\end{eqnarray}
where $\alpha$ is the elecromagnetic fine-structure constant. 
In the mass insertion approximation, the A-term contribution is that
\begin{eqnarray}
\begin{split}
\label{MIamplL}
A^{ij}_L
\simeq& \frac{\alpha_1}{4\pi}~\frac{\left(\delta^{RL}_{\ell}\right)_{ij}}{m_{\tilde \ell}^2}~
\left(\frac{M_1}{m_{\ell_i}}\right) \times 2~f_{2n}(x_1)~,
\\
A^{ij}_R
\simeq&
\frac{\alpha_{1}}{4\pi}
\frac{\left(\delta^{LR}_{\ell}\right)_{ij}}{m_{\tilde\ell}^{2}}~
\left(\frac{M_1}{m_{\ell_i}}\right)\times 2~f_{2n}(x_1)~,
\end{split}
\end{eqnarray}
where 
$x_{1}=M_{1}^2/m_{\tilde \ell}^2$ and  $\alpha_1=g_1^2/4\pi$.
Mass parameters 
$M_1$ and  $m_{\ell_i}$ are  the  $U(1)_Y$ gaugino mass  
and  the charged lepton mass, respectively. 
The  loop function 
$f_{2n}(x)$ is  given explicitly as follows:
\begin{eqnarray}
f_{2n}(x) = \frac{-5x^2+4x+1+2x(x+2)\log x}{4(1-x)^4}~. 
\end{eqnarray}
The mass insertion parameters
$\delta^{RL}_{\ell}$ and $\delta^{LR}_{\ell}$ 
are given in Eq.\,(\ref{massinsertion}).
The contributions of $\delta^{LL}_{\ell}$ and $\delta^{RR}_{\ell}$ 
are neglected because the off diagonal components vanish
as discussed in  Eq.\,(\ref{massLe}).
In SUSY models, the branching ratio of $\ell_i \to 3\ell_j$ and conversion rate of $\mu N \to e N$ also can be 
related as
\begin{align}
\frac{{\rm BR}(\ell_{i}\rightarrow  \ell_{j} \ell_{k} \bar{\ell}_{k})}{{\rm BR}(\ell_{i}\rightarrow \ell_{j}\gamma)} &\simeq
\frac{\alpha}{3 \pi} \left( 2 \log\frac{m_{\ell_i}}{m_{\ell_k}} - 3\right)\,. \\
\frac{{\rm CR}(\mu N\to e N)}{{\rm BR}(\ell_{i}\rightarrow \ell_{j}\gamma)} &\simeq \alpha.
\end{align}

In numerical calculations of the  $\mu\rightarrow e + \gamma$ ratio, 
we take a sample parameter set  to be consistent with
the observed lepton masses and flavor mixing angles
in the model of Table 1   \cite{Okada:2019uoy,Okada:2020ukr,Okada:2020brs} as follows:
\begin{equation}
{\bf A}\,: \ \  \tau=-0.0796 + 1.0065  \, i \,, \qquad  
\alpha_e/\gamma_e=6.82\times 10^{-2}\,, \qquad 
\beta_e/\gamma_e=1.02\times 10^{-3}\,,
\label{tau0}
\end{equation}
which is referred as the parameter set {\bf A}.
This model favors the modulus  $\tau$ being close to $i$,
where important physics such as $CP$
 and the hierarchy of fermion masses are revealed 
\cite{Novichkov:2019sqv,Kobayashi:2020uaj,Novichkov:2021evw,Feruglio:2021dte}.
In this sample,   $\tan\beta=5$ is taken in order to fit 
the lepton masses at  hight energy scale \cite{Okada:2019uoy,Okada:2020ukr,Okada:2020brs}.
On the other hand, 
the SUSY mass parameters are the gaugino mass $M_1$ and  the averaged slepton mass $m_{\tilde\ell}$, which are  low energy
observables at the EW scale. The SUSY breaking parameter $F^\tau$
is expected to be the same order compared with $m_{\tilde\ell}$.

In order to see  the SUSY mass scale dependence for 
the $\mu\rightarrow e + \gamma$ branching ratio,
we show  them (red curves)  by taking the  averaged  mass scale $m_0\equiv m_{\tilde\ell}=F^\tau$ 
with $M_1=3$\,TeV\,(solid curve) and $5$\,TeV\,(dashed curve) 
for simplicity in Fig.\,1.
The SUSY mass scale $m_0$ should be larger than around $8$\,TeV
to be consistent with the observed upper bound (black line). 
The predicted LFV branching ratio can be examined by future experiment such as MEG-II \cite{Baldini:2018nnn} 
(orange line) up to $m_0 \simeq 12$ TeV.
The magnitudes of  the mass insertion parameters are proportional to
 $F^\tau$.  Those  are given as
\begin{equation}
|(\delta^{RL}_{\ell})_{\mu e}|\simeq 2.1\times 10^{-5} \left (\frac{F^\tau}{10\,{\rm TeV}}\right )\,, \qquad \quad
|(\delta^{LR}_{\ell})_{\mu e}|\simeq 9.7\times 10^{-8}
\left (\frac{F^\tau}{10\,{\rm TeV}}\right )\,.
\end{equation}
Therefore, the amplitude $|A_L^{\mu e}|$ is much larger than $|A_R^{\mu e}|$.

\begin{figure}[h]
	\begin{tabular}{ccc}
		\begin{minipage}{0.48\hsize}
			\includegraphics[bb=0 0 200 160,width=\linewidth]{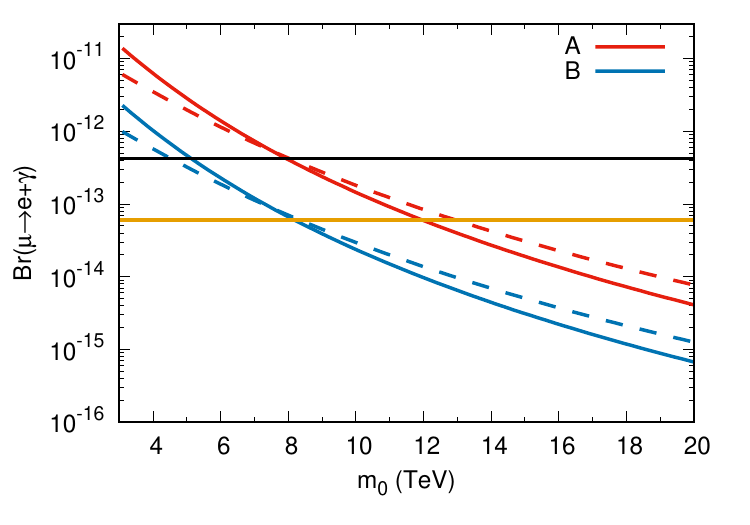} \vspace{-7mm}
	\caption{Branching ratio of $\mu \rightarrow e + \gamma$  for the SUSY mass scale $m_0\equiv m_{\tilde\ell}=F^\tau$
		 for parameter sets {\bf A}(red:\,$\tau=-0.0796 + 1.0065\,i$) and {\bf B}(blue:\,$\tau=0.48151 + 1.30262 \,i$), respectively.
		 	The solid and dashed curves correspond to 	
		 	$M_1=3$\,TeV and $5$\,TeV, respectively.
	The horizontal black and orange lines are the experimental upper bound and future expected bound. }
			\label{}
		\end{minipage}
		\hskip 0.7 cm
		\begin{minipage}{0.5\hsize}
			\includegraphics[bb=0 0 220 155,width=\linewidth]{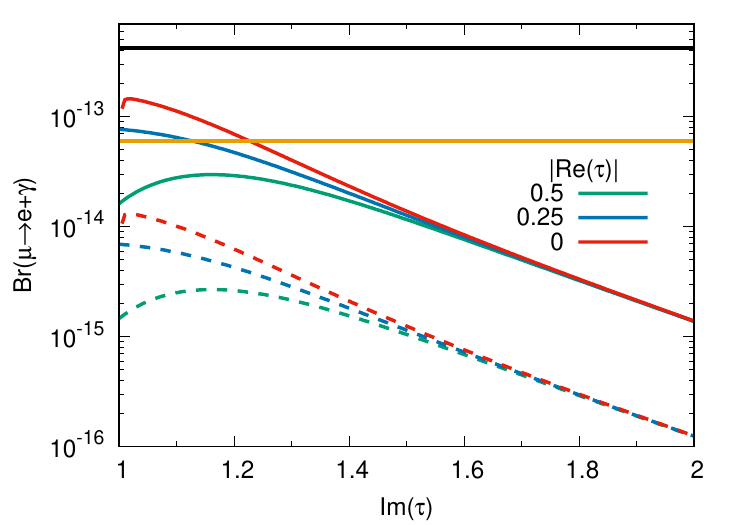}
			\caption{Branching ratio of $\mu \rightarrow e + \gamma$  for ${\rm Im}\, \tau$.
			The solid and dashed curves correspond to 	$F^\tau=m_{\tilde\ell}$ and  $F^\tau=M_1$ with 
				$m_{\tilde\ell}=10$\,TeV and  $M_1=3$\,TeV, respectively.
			Red, blue and green curves denote
				$|{\rm Re}\, \tau|=0,\, 0.25$ and $ 0.5$ in the fundamental region $SL(2,\mathbb{Z})$, respectively.
				The horizontal black and orange lines are the experimental upper bound and future expected bound.}
			\label{}
		\end{minipage}
	\end{tabular}
\end{figure}

In this calculation, the model of lepton mass matrices
is fixed as seen in Table 1.
There are variant neutrino mass matrices in modular flavor $A_4$ model,
where the charged lepton mass matrix is the same one in Eq.\,(\ref{eq:CL})
\cite{Kobayashi:2018scp,Asaka:2019vev,Kobayashi:2019gtp}.
In those models,  the A-term of the neutrino sector
appears due to right-handed neutrinos. 
However, this contribution is suppressed as far as
the seesaw mechanism works at high energy.

A simple alternative model is presented in Table \ref{tb:fields},
where neutrino masses are generated via seesaw mechanism
by introducing the right-handed neutrino $\nu^c$.
\begin{table}[h]
	\centering
	\begin{tabular}{|c||c|c|c|c|c|c|} \hline 	\rule[14pt]{0pt}{1pt}
		&$L$&$(e^c,\mu^c,\tau^c)$&$\nu^c$&$H_u$&$H_d$&$Y_{\bf 3}^{(2)}$\\ \hline \hline 
		\rule[14pt]{0pt}{0pt}
		$SU(2)$&$\bf 2$&$\bf 1$&$\bf 1$&$\bf 2$&$\bf2$&$\bf 1$\\
		$A_4$&$\bf 3$& \bf (1,\ 1$''$,\ 1$'$)&$\bf 3$&$\bf 1$&$\bf 1$&$\bf 3$\\
		$k$&$-1$&$(-1,\ -1,\ -1)$&$-1$&0&0&$2$ \\ \hline
	\end{tabular}
	\caption{Representations of $SU(2)$, $A_4$, and the modular weight in the type I seesaw model.}
	\label{tb:fields}
\end{table}
A sample parameter set, referred the parameter set {\bf B}, to be consistent with the observed lepton masses and flavor mixing angles
as follows 	\cite{Kobayashi:2018scp,Asaka:2019vev}:
\begin{equation}
{\bf B}\, : \ \  \tau=0.48151 + 1.30262 \,i \,, \qquad  
\alpha_e/\gamma_e=2.03\times 10^{2}\,, \qquad 
\beta_e/\gamma_e=3.30\times 10^{3}\,,  
\end{equation}
where $\tan\beta=10$ is taken.
The magnitude of modulus $\tau$ is larger than
the one in Eq.\,(\ref{tau0}).

We also show  the branching ratio (blue) versus the SUSY mass scale $m_0$ in Fig.\,1.
In this case, the SUSY mass scale $m_0$ should be larger than around $5$\,TeV
to be consistent with the observed upper bound. 
The predicted LFV branching ratio can be examined by future experiment (orange line) 
up to $m_0 \simeq 8$ TeV.
The magnitudes of  the mass insertion parameters  are given as
\begin{equation}
|(\delta^{RL}_{\ell})_{\mu e}|\simeq 8.4\times 10^{-6} \left (\frac{F^\tau}{10\,{\rm TeV}}\right )\,, \qquad \quad
|(\delta^{LR}_{\ell})_{\mu e}|\simeq 3.7\times 10^{-8}
\left (\frac{F^\tau}{10\,{\rm TeV}}\right )\,.
\end{equation}
The amplitude $|A_L^{\mu e}|$ is much larger than $|A_R^{\mu e}|$
 as well as the case of the parameter set {\bf A}.


The  predicted branching ratio
apparently decreases as the magnitude of  $\tau$ increase
as seen in Fig.\,1.
In order to see the ${\rm Im}\,\tau$ dependence of the branching ratio,
we show the branching ratio versus ${\rm Im}\,\tau$ in Fig.2,
where solid  and dashed curves correspond to
$F^\tau=m_{\tilde\ell}$ and  $F^\tau=M_1$ with 
$m_{\tilde\ell}=10$ TeV and $M_1=3$ TeV, respectively.
We choose  $|{\rm Re}\,\tau | =0,\,0.25,\,0.5$
in the fundamental region $SL(2,\mathbb{Z})$.
  For each $\tau$ of Fig.\,2, 
    we do not take into account lepton mixing angles 
    consistent with observed ones since they depend on the 
     model of the neutrino mass matrix.
    The charged lepton mass matrix of  Eq.\,(\ref{eq:CL}) is completely determined   by inputting observed charged lepton masses
    if   $\tau$ is fixed.
 Therefore, we can see the   ${\rm Im}\,\tau$ dependence
of the branching ratio for each ${\rm Re}\,\tau$ as seen in Fig.\,2.
The branching ratio depends on both  
${\rm Im}\,\tau$ and ${\rm Re}\,\tau$ significantly below
${\rm Im}\,\tau\simeq 1.4$.
Thus, the branching ratio changes more than one order
depending on $\tau$.

Figures 3 and 4 show the SUSY mass scale dependence for the $\mu \to 3e$ branching ratio and $\mu N \to e N$ 
conversion rate in the same parameter sets in Fig.~1. We can see that the predicted branching ratio and conversion rate 
are enough below the current experimental bound for $m_0 > 3$ TeV. Future experiments will explore these predictions 
at the level of $10^{-16}$ \cite{Blondel:2013ia,Carey:2008zz,Cui:2009zz,Wong:2015fzj}, which corresponds to $m_0 \simeq 10$\,--\,$16$ TeV in $\mu \to 3e$ decay and $11$\,--\,$17$ TeV in 
$\mu \to e$ conversion. 
\begin{figure}[h]
	\begin{tabular}{ccc}
		\begin{minipage}{0.48\hsize}
			\includegraphics[bb=0 0 205 155,width=\linewidth]{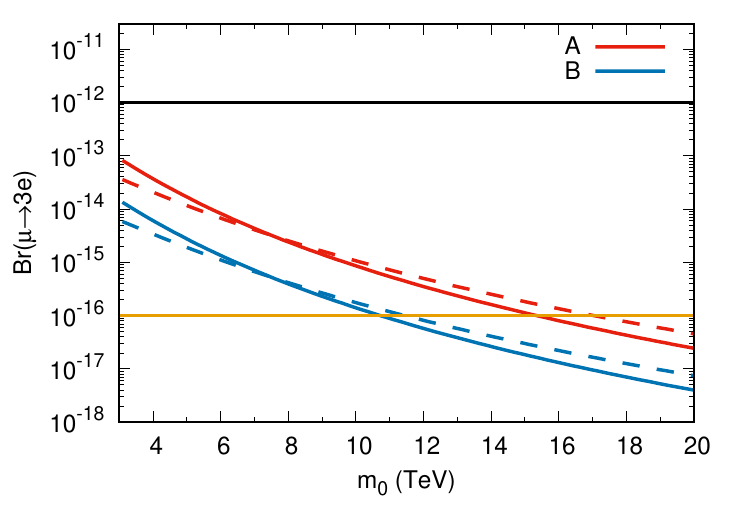}
	\caption{Branching ratio of $\mu \rightarrow 3e $  for the SUSY mass scale $m_0\equiv m_{\tilde\ell}=F^\tau$
			 with the same parameter sets in Figure 1.
		The horizontal black and orange lines are the experimental upper bound and future expected bound. }
			\label{}
		\end{minipage}
		\hskip 0.7 cm
		\begin{minipage}{0.5\hsize}
			\includegraphics[bb=0 0 220 155,width=\linewidth]{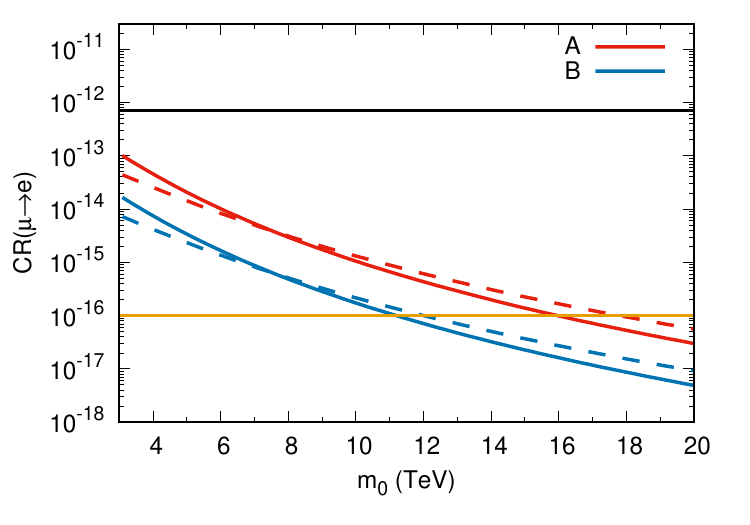}
			\caption{Conversion rate of $\mu N to e N$  for the SUSY mass scale $m_0\equiv m_{\tilde\ell}=F^\tau$
			 with the same parameter sets in Figure 1.
				The horizontal black and orange lines are the experimental upper bound and future expected bound.}
			\label{}
		\end{minipage}
	\end{tabular}
\end{figure}

In conclusion, the current experimental search for the 
$\mu\to e +\gamma$ decay provides a clue of the SUSY particles at the  $5$\,--\,$10$\,TeV scale in the modular flavor  models. 
The predictions of modular flavor models will be examined in future experimental searches up to $8$\,--\,$17$ TeV scale.
Lastly, we comment on the Higgs mass and SUSY particle masses. There exist many 
parameters to determine the soft masses unless SUSY breaking mechanism is specified. Adjusting those parameters, 
it will be possible to obtain the Higgs and SUSY spectrum which is consistent with the current LHC bounds. 
However, such analyses are beyond the scope of this paper and left for future studies.


We can also calculate the branching ratios of tauon decays, 
$\tau\rightarrow e + \gamma$ and $\tau\rightarrow \mu + \gamma$.
Their predicted branching ratios are
at most ${\cal O}(10^{-15})$, which are  much  below the current experimental  bounds. 
The present and future bounds on these processes are summarized 
in Table \ref{tb:lfv} \cite{TheMEG:2016wtm,Zyla:2020zbs} and 
\cite{Baldini:2018nnn,Blondel:2013ia,Carey:2008zz, Cui:2009zz,Wong:2015fzj}.
\begin{table}[h]
	\addtolength{\arraycolsep}{3pt}
	\renewcommand{\arraystretch}{1.3}
	\centering
	\begin{tabular}{|c||c|c|c|c|c|}
		\hline
		Processes & BR($\mu \to e \gamma$) & BR($\mu \to 3 e $) & CR($\mu N \to e N$) 
		&BR($\tau \to e \gamma$) &BR($\tau \to \mu \gamma$) \\
		\hline
		~~Current bound~~
		& $4.2~ \times~ 10^{-13}$ & $1.0 \times 10^{-12} $ & $7.0 \times 10^{-13}$ 
		& $3.3~ \times~ 10^{-8}$ & $4.4~ \times~ 10^{-8}$\\
		\hline
		Future bound
		& $6 \times 10^{-14}$ & $10^{-16}$ & $10^{-16}$
		& --- & --- \\
		\hline
	\end{tabular}
	\caption{\small
		Present and future upper bounds of the lepton flavor violation for each process 
		\cite{TheMEG:2016wtm,Zyla:2020zbs} and  \cite{Baldini:2018nnn,Blondel:2013ia,Carey:2008zz,Cui:2009zz,Wong:2015fzj}.}
	\label{tb:lfv}
\end{table}

\section{Summary}

We have studied  the soft SUSY breaking terms due to the modulus F-term in the modular flavor models of leptons.  It is found that the soft SUSY breaking terms
are constrained by the modular forms, and the specific pattern of soft SUSY breaking terms appears.

Those  phenomenological implications have been  discussed  in such as the lepton flavor violation,  
$\mu \rightarrow e + \gamma$ and $\mu \to 3e$ decays, and $\mu \to e$ conversion.
In order to examine numerically, 
 parameter sets {\bf A}  and {\bf B}
 are adopted in  two modular flavor $A_4$ models. 
The SUSY mass scale is significantly constrained by inputting the observed upper bound of the $\mu \rightarrow e + \gamma$ decay.
The SUSY mass scale  is  larger than around $8$\,TeV
and $5$\,TeV for parameter sets {\bf A}  and {\bf B}, respectively.
Therefore, the  current experimental upper bound for the $\mu \to e + \gamma$ decay
corresponds to  the new physics of  the SUSY particles at the  $5$\,--\,$10$\,TeV scale  in the modular flavor $A_4$ models.
The predicted branching ratio and conversion rate will be examined by future experiments for the SUSY scale 
up to $8$\,--\,$17$ TeV.
The branching ratio depends on $\tau$ significantly.
It decreases of one order  at the large  
 ${\rm Im}\,\tau$.
We have  also calculated the branching ratios of tauon decays, 
$\tau\rightarrow e + \gamma$ and $\tau\rightarrow \mu + \gamma$.
Their predicted branching ratios are
at most ${\cal O}(10^{-15})$, which are  much  below the current experimental  bounds. 

It is important to perform similar analyses in other modular flavor models.
These specific patterns of soft SUSY breaking terms
 of the modular flavor models
can be tested in the future experiments of the lepton flavor violations.

\vspace{1cm}
\noindent
{\bf Acknowledgement}

This work is supported by  MEXT KAKENHI Grant Number JP19H04605 (TK),~JP18H05543 (TS) and 
JSPS KAKENHI Grant Number JP18H01210 (TS),~JP18K03651 (TS).

\newpage
\noindent
{\LARGE \bf Appendix}
\appendix

\section{Tensor product of  $A_4$ group}
We take the generators of $A_4$ group for the triplet as follows:
\begin{align}
\begin{aligned}
S=\frac{1}{3}
\begin{pmatrix}
-1 & 2 & 2 \\
2 &-1 & 2 \\
2 & 2 &-1
\end{pmatrix},
\end{aligned}
\qquad 
\begin{aligned}
T=
\begin{pmatrix}
1 & 0& 0 \\
0 &\omega& 0 \\
0 & 0 & \omega^2
\end{pmatrix}, 
\end{aligned}
\end{align}
where $\omega=e^{i\frac{2}{3}\pi}$ for a triplet.
In this base,
the multiplication rule is
\begin{align}
\begin{pmatrix}
a_1\\
a_2\\
a_3
\end{pmatrix}_{\bf 3}
\otimes 
\begin{pmatrix}
b_1\\
b_2\\
b_3
\end{pmatrix}_{\bf 3}
&=\left (a_1b_1+a_2b_3+a_3b_2\right )_{\bf 1} 
\oplus \left (a_3b_3+a_1b_2+a_2b_1\right )_{{\bf 1}'} \nonumber \\
& \oplus \left (a_2b_2+a_1b_3+a_3b_1\right )_{{\bf 1}''} \nonumber \\
&\oplus \frac13
\begin{pmatrix}
2a_1b_1-a_2b_3-a_3b_2 \\
2a_3b_3-a_1b_2-a_2b_1 \\
2a_2b_2-a_1b_3-a_3b_1
\end{pmatrix}_{{\bf 3}}
\oplus \frac12
\begin{pmatrix}
a_2b_3-a_3b_2 \\
a_1b_2-a_2b_1 \\
a_3b_1-a_1b_3
\end{pmatrix}_{{\bf 3}\  } \ , \nonumber \\
\nonumber \\
{\bf 1} \otimes {\bf 1} = {\bf 1} \ , \qquad &
{\bf 1'} \otimes {\bf 1'} = {\bf 1''} \ , \qquad
{\bf 1''} \otimes {\bf 1''} = {\bf 1'} \ , \qquad
{\bf 1'} \otimes {\bf 1''} = {\bf 1} \  ,
\end{align}
where
\begin{align}
T({\bf 1')}=\omega\,,\qquad T({\bf 1''})=\omega^2. 
\end{align}
More details are shown in the review~\cite{Ishimori:2010au,Ishimori:2012zz}.

By using above tensor products of  the modular forms  with weight 2,
${\bf Y^{\rm (2)}_3}(\tau)$,
the  modular forms of the  higher weight, $k$, are obtained.
For weight 4, that is $k=4$, there are  five modular forms
by the tensor product of  $\bf 3\otimes 3$ as:
\begin{align}
&\begin{aligned}
{\bf Y^{\rm (4)}_1}(\tau)=Y_1(\tau)^2+2 Y_2(\tau) Y_3(\tau) \, , \qquad\quad\ \
{\bf Y^{\rm (4)}_{1'}}(\tau)=Y_3(\tau)^2+2 Y_1(\tau) Y_2(\tau) \, , 
\end{aligned}\nonumber \\
\nonumber \\
&\begin{aligned} 
{\bf Y^{\rm (4)}_{1''}}(\tau)=Y_2(\tau)^2+2 Y_1(\tau) Y_3(\tau)=0 \, , \qquad
{\bf Y^{\rm (4)}_{3}}(\tau)=
\begin{pmatrix}
Y_1^{(4)}(\tau)  \\
Y_2^{(4)}(\tau) \\
Y_3^{(4)}(\tau)
\end{pmatrix}
=
\begin{pmatrix}
Y_1(\tau)^2-Y_2(\tau) Y_3(\tau)  \\
Y_3(\tau)^2 -Y_1(\tau) Y_2(\tau) \\
Y_2(\tau)^2-Y_1(\tau) Y_3(\tau)
\end{pmatrix}\, , 
\end{aligned}
\label{weight4}
\end{align}
where ${\bf Y^{\rm (4)}_{1''}}(\tau)$ vanishes due to the constraint of
Eq.\,(\ref{condition}).



\end{document}